\documentclass[journal,times,mathptm, oneside]{IEEEtran}%

\usepackage{multirow}  \usepackage{cite}
\usepackage{graphicx,subfigure}\usepackage{amsmath} \usepackage{amssymb} \usepackage{amsthm}\usepackage{color,array}
\usepackage{etex,etoolbox} \usepackage{float}
\usepackage{color}\usepackage{graphicx}
\usepackage{caption}\usepackage{xcolor}
\usepackage{colortbl}
\usepackage{soul}
\usepackage{algorithm}
\usepackage{soul,color,bm}
\usepackage{setspace}

\usepackage{algorithmic}   \usepackage{bbm}
\usepackage[T1]{fontenc}

\makeatletter
\providecommand{\@fourthoffour}[4]{#4}
\def\fixstatement#1{%
  \AtEndEnvironment{#1}{%
    \xdef\pat@label{\expandafter\expandafter\expandafter
      \@fourthoffour\csname#1\endcsname\space\@currentlabel}}}

\long\def\proofatend#1\endproofatend{%
  \edef\next{ \Alph{proofcount}. Proof of \pat@label \noexpand\begin{proof}[Proof]}%
  \toks\numexpr\prooftoks+\value{proofcount}\relax=\expandafter{\next#1\end{proof}}
  \stepcounter{proofcount}}

\def\printproofs{%
  \count@=\z@
  \loop
    \the\toks\numexpr\prooftoks+\count@\relax
    \ifnum\count@<\value{proofcount}%
    \advance\count@\@ne
  \repeat}

\makeatother

\fixstatement{thm}
\fixstatement{lem}

\begin{document}


\title{LLM-Empowered Resource Allocation in \\Wireless Communications Systems}
\author{Woongsup Lee and Jeonghun Park

\thanks{W. Lee is with the Graduate School of Information, Yonsei University, South Korea. J. Park is with the School of Electrical and Electronic Engineering, Yonsei University, South Korea. (E-mail: {\texttt{\{woongsup.lee, jhpark\}@yonsei.ac.kr}})}}

\maketitle



\begin{abstract}

The recent success of large language models (LLMs) has spurred their application in various fields. In particular, there have been efforts to integrate LLMs into various aspects of wireless communication systems. The use of LLMs in wireless communication systems has the potential to realize artificial general intelligence (AGI)-enabled wireless networks. In this paper, we investigate an LLM-based resource allocation scheme for wireless communication systems. Specifically, we formulate a simple resource allocation problem involving two transmit pairs and develop an LLM-based resource allocation approach that aims to maximize either energy efficiency or spectral efficiency. Additionally, we consider the joint use of low-complexity resource allocation techniques to compensate for the reliability shortcomings of the LLM-based scheme. After confirming the applicability and feasibility of LLM-based resource allocation, we address several key technical challenges that remain in applying LLMs in practice.

\end{abstract}

\section{Introduction} 


In wireless communication systems, the allocation of resources such as transmit power, bandwidth, or beamforming is of utmost importance because the openness nature of wireless medium causes interference to neighboring nodes \cite{Xu2021}. Recently, the number of transmitting nodes has increased, while the required communication objectives have become more diverse and stringent. At the same time, more complicated system models and stringent constraints, such as extremely low computation time, are being considered. These factors complicate resource allocation, spurring extensive research in this field.

Since finding the optimal resource allocation strategy is typically formulated as an optimization problem, analytical optimization frameworks such as convex optimization have been widely applied \cite{Liu2024}. However, given that the Shannon capacity formula is nonconvex with respect to transmit power and many control variables, such as channel selection, are discrete, finding the optimal resource allocation is challenging. To address these issues, novel mathematical approaches, such as game-theoretic frameworks or convex relaxation techniques, have been developed \cite{Liu2024}. Despite these advances, most analytical approaches face limitations in efficiently managing large numbers of nodes at low computational times and struggle to adapt to dynamically changing wireless environments.

The deep learning (DL)-based resource allocation has received considerable attention to address the above-mentioned challenges \cite{Sun2018, Shi2023}. In DL-based schemes, the optimal resource allocation strategy is approximated using specially structured deep neural network (DNN) models. Since DNNs primarily perform simple matrix operations, once properly trained, they can achieve near-optimal performance with significantly low computation times, on the order of a few milliseconds. Despite these advantages, the design and training of DNNs is task-specific, requiring unique structures and training for each resource allocation task. This limits the applicability of the DL-based approach. 

Very recently, the large language model (LLM), which is the foundational model built on natural language understanding, has achieved significant success in various domains, exemplified by technologies such as OpenAI's ChatGPT and Google's Bard. These models are considered game changers in the next wave of artificial intelligence (AI). LLM-based technology can also open new dimensions in wireless network design and operation, potentially leading to much higher performance \cite{Bariah2024}. In particular, the multimodal data relevant to wireless communication systems, such as radio frequency signals and visual representations of the wireless environment, can be properly exploited in LLM-based approaches, which enables situational and temporal awareness and predictability, thus facilitating proactive control of wireless communication systems.

Thanks to the human-like reasoning and inference capabilities of LLM, efforts have been made to use them to solve mathematical problems, and it has been demonstrated that LLMs can indeed be effective in this domain. Consequently, it is promising to employ the LLMs to address various optimization problems in wireless communication systems, such as finding optimal resource allocation strategies. Compared to the conventional DL-based approaches, LLM-based approaches offer the significant advantage of providing reasonable outputs without the need for designing task-specific model design and training, since LLMs are already pre-trained on extensive datasets. Thus, the LLM can be considered as a general-purpose solver capable of finding optimal solutions to a wide range of optimization problems, making it easily adaptable to different environments.

In this article, we investigate the feasibility of LLM-based resource allocation in wireless communication systems. We begin by describing the basic principles of LLMs and exploring how they can be leveraged to determine resource allocation strategies. Next, we provide an illustrative example of a simple resource allocation problem that aims to maximize either spectral efficiency (SE) or energy efficiency (EE), and design an LLM-based resource allocation framework. 
In addition, we also propose a hybrid resource allocation strategy that combines LLM-based resource allocation with conventional resource allocation techniques. Through simulations, we validate the feasibility of LLM-based resource allocation and evaluate its performance. Finally, we outline the key research challenges associated with applying LLMs to resource allocation and present some key conclusions. 
\begin{figure*}[t]
	\centerline{\includegraphics[width=17.2cm]{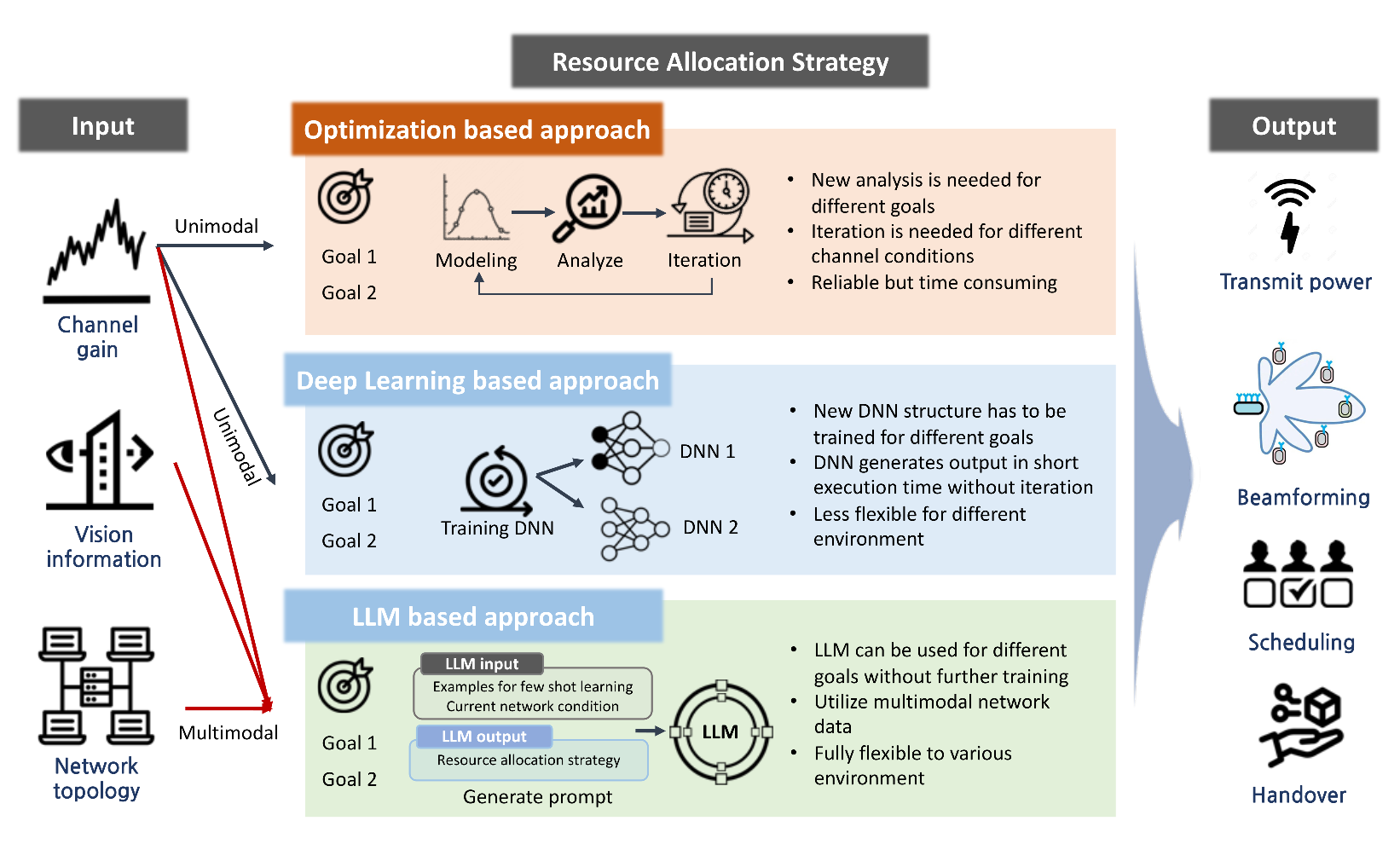}}
	\caption{Comparison between different approaches for resource allocation.}
	\label{fig_e-to-e-learn} 
\end{figure*}

\section{Basic Principle of LLM and Its Application in Resource Allocation} 

In this section, we describe the basic principles of LLM. Then, we turn our attention to the feasibility and benefits of using LLMs for resource allocation.

\subsection{Principle of LLM}

A DNN, trained on extremely large amounts of unlabeled data using self-supervision techniques to adapt to a wide range of tasks, is referred to as a foundation model. The foundation model is designed to be highly generalizable and can be fine-tuned for specific applications, making it a versatile and powerful base for a wide range of tasks. Notably, the foundation models typically have deep architectures with many layers and a large number of parameters (e.g., 175 billion parameters for GPT-3), allowing them to capture complex relationships in data and perform complicated tasks that require logical reasoning and understanding of the data. The LLM is a representative example of a foundation model, trained on large amounts of text data. Accordingly, the LLM generates text and performs text-based tasks based on given prompts, which are natural language descriptions of the tasks. 

LLM is primarily based on the Transformer architecture, which has become the mainstream framework for modern language models. The Transformer contains two main innovations: the attention mechanism and positional encoding \cite{Vaswani2017}. The attention mechanism computes the relevance of each word in a sentence relative to other words, allowing the model to capture dependencies between words regardless of their distance in the text. This mechanism is enhanced by multi-head attention, which enables the model to focus on different parts of the input sequence simultaneously, capturing different contextual aspects. In addition, positional encoding is added to the input embeddings to encode information about the position of each word in the sequence, ensuring that the order and relative positioning of words are taken into account. 

Most well-known LLMs are pre-trained on billions to trillions of tokens over millions of iterations. Consequently, fine-tuning with task-specific labeled datasets is often employed to further adapt the LLM to specific tasks \cite{Kojima2022}. However, given the enormous number of trainable parameters, the computational overhead for fine-tuning can be substantial and costly. Given the reasoning capabilities of the LLM, few-shot or zero-shot learning approaches, where only a few concrete examples of a task are provided, or no examples at all, can be effective. Despite the minimal input, the LLM's inductive reasoning capabilities allow it to produce reasonable results. For example, simply adding ``Let's think step by step" to the prompt can guide the LLM to follow a human-like reasoning process, leading to better results \cite{Kojima2022}.

Given their importance, leading AI companies have developed their own LLMs. Among the best known are GPT, developed by OpenAI; and LLaMA, developed by Meta \cite{Minaee2024}.

\begin{itemize}

\item \textbf{GPT :} The GPT, or Generative Pre-trained Transformers, are decoder-only Transformer-based language models developed by OpenAI. The early models, such as GPT-1 and GPT-2, are open source, while more recent models, such as GPT-3 and GPT-4, are closed source. Although the basic GPT model is trained on general language data, specific versions such as InstructGPT and ChatGPT are trained for specific tasks. These models use reinforcement learning from human feedback to follow instructions in a more human-like manner \cite{Minaee2024}.
\vspace{1mm}

\item \textbf{LLaMA :} LLaMA, or Large Language Model Meta AI, is a collection of basic language models published by Meta. A notable aspect of LLaMA is that, unlike GPT, it is open source, meaning that model weights are made available. In addition, the size of LLaMA models is relatively small compared to GPT-based models, ranging from 7 billion to 65 billion parameters. As a result, LLaMA has been widely used in the research community and in the development of task-specific LLMs for mission-critical applications. 

\end{itemize}

\subsection{Conventional Resource Allocation Strategies}

In resource allocation for wireless communication systems, parameters such as transmit power are adjusted to optimize network performance, such as maximizing SE, while satisfying constraints. Finding the optimal strategy is typically formulated as an optimization problem, which can be addressed by using either conventional optimization-based approaches or DL-based approaches as follows \cite{Lee2020}:

\begin{itemize}

\item \textbf{Optimization-based approach :} The formulated optimization problem is tackled analytically. In general, the optimization problem is often complicated and hard to solve, involving aspects such as integer-valued control parameters and non-convex functions. As a result, a variety of mathematical approaches are typically used to transform and simplify the problem, to make it more tractable. The optimal solution is then often obtained through iterative methods.
\vspace{1mm}

\item \textbf{DL-based approach :} The optimal resource allocation is approximated using the output of a specially designed DNN, which provides the resource allocation strategy based on the current wireless network conditions. To achieve this, the DNN must be trained, which can be done either through supervised learning exploiting optimal resource allocation as labeled data, or through unsupervised learning without such labeled data. Although training the DNN can take a considerable amount of time, inference from the trained DNN can typically be performed in the order of a few milliseconds.

\end{itemize}

Although resource allocation has been extensively investigated in the literature, conventional methods face significant limitations, especially with respect to multimodality and flexibility. In particular, resource allocation in future wireless systems should be capable of handling multimodal data, such as geographic information. By effectively utilizing such multimodal data, future channel conditions can be predicted, enabling proactive resource allocation, which is crucial for achieving ultra-low latency. In addition, resource allocation should be adaptable to various communication objectives and topologies to accommodate diverse scenarios. Conventional approaches often lack these capabilities. While the DL-based approaches can be trained to flexibly adapt to changing environments, they require specially designed DNNs, and the training process can be time-consuming.
\begin{figure*}[t]
	\centerline{\includegraphics[width=17.5cm]{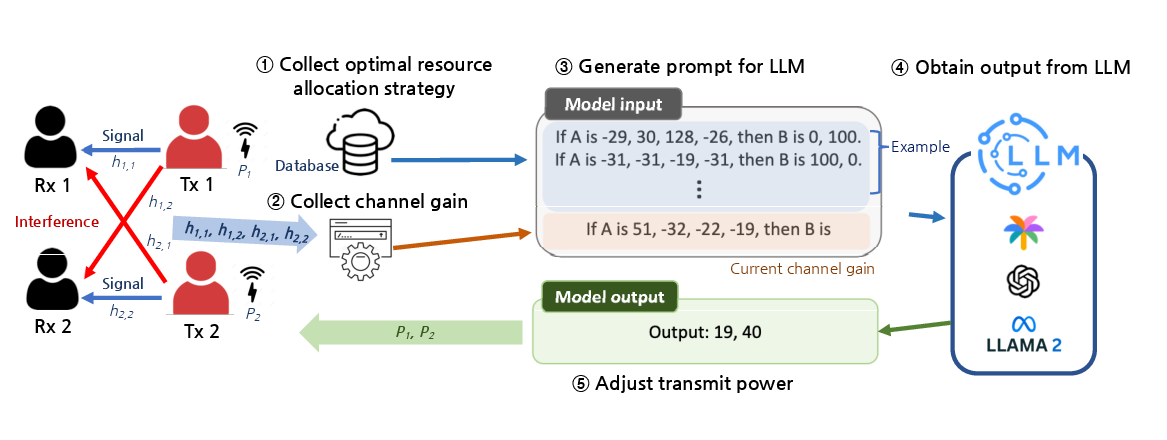}}
	\caption{Considered system model and procedure for LLM-based resource allocation.}
	\label{fig_dnn_model}
\end{figure*}

\subsection{Resource Allocation Using LLM}

LLM-based resource allocation presents a promising solution to overcome the drawbacks of conventional schemes. Notably, in this approach, resource allocation can be determined by the LLM on a few-shot learning basis, eliminating the need for retraining of the LLM for specific scenarios. In essence, the LLM acts as a general optimizer, addressing a wide range of resource allocation problems. The benefits of using LLMs in resource allocation can be summarized as follows:

\begin{itemize}

\item \textbf{Multimodality :} LLM-based scheme can handle various multimodal data. By effectively using such multimodal data in wireless communication systems, more efficient and proactive radio resource allocation can be achieved.
\vspace{1mm}

\item \textbf{Flexibility :} Unlike DL-based schemes that require extensive training, LLMs can operate without task-specific training. Consequently, LLMs can easily adapt to a wide range of objectives and topologies.

\end{itemize}

To effectively use LLMs for resource allocation, it is necessary to validate that they can solve mathematical problems. There have been several attempts to use LLMs to solve linguistically described mathematical problems, such as the GSM8K dataset. Although LLMs are primarily designed for linguistic tasks and may not be well suited for rigorous mathematical computations, recent studies indicate that LLMs can indeed solve mathematical problems when properly prompted. For example, the authors of \cite{Zhong2024} showed that LLMs can accurately solve mathematical problems by formulating prompts, achieving an accuracy of 97.1\% in a zero-shot setting. Moreover, in \cite{Besta2024}, it was confirmed that the mathematical capabilities of LLMs could be enhanced through sequential reasoning techniques such as Chain of Thoughts, Tree of Thoughts, and Graph of Thoughts. In \cite{Joy2023}, the use of external math solvers was explored to address the limitations of LLM in precise algebraic computation. In addition, the authors of \cite{Chengrun2023} demonstrated that LLMs could be used to solve optimization problems, such as the linear regression and the traveling salesman problem, through iterative optimization processes. Despite these previous works, there have been no attempts to apply LLMs to resource allocation in wireless communication systems.

\section{LLM-based Resource Allocation} 

In this section, we illustrate the feasibility of using LLMs for resource allocation. To this end, we first describe a simplified resource allocation problem. We then explain how LLMs can be utilized for the resource allocation and present a scheme designed to compensate for the shortcomings of a purely LLM-based approach. Finally, the feasibility of LLM-based resource allocation is evaluated through simulation.

\subsection{System Model and Considered Resource Allocation}

We consider a resource allocation strategy for two transmit pairs, which are randomly distributed. Specifically, we assume that two transmitters send data to their respective receivers over the same channel. Let $h_{i, j}$ denote the channel gain between transmitter $i$ and receiver $j$, where $i, j \in \{1, 2\}$, and $P_{i}$ denotes the transmit power of the transmitter $i$, where the transmit power for a transmitter must not exceed the maximum transmit power, $P_{\textrm{T}}$.

We consider two resource allocation strategies to maximize either total SE or EE by properly adjusting the transmit power. The achievable SE of the transmit pair $i$, denoted as $\textrm{SE}_{i}$, can be expressed as $\log_{2} \left(1+\frac{ h_{i,i} P_{i}}{N_{0}W + \sum_{l\in \{1, 2\} \setminus \{i\} h_{l,i} P_{l}}}\right) $, where $N_{0}$ and $W$ are noise power density and bandwidth, respectively. Then the objective for maximizing the total SE can be formulated as $\sum_{i=1}^{2} \textrm{SE}_{i}$. Conversely, the objective for maximizing total EE can be formulated as $\sum_{i=1}^{2} \frac{\textrm{SE}_{i}}{P_{i} + P_{\textrm{C}}}$, where $P_{\textrm{C}}$ is the constant circuit power of the transmitter. Although we do not explicitly show it in the paper, the strategy for transmit power control behaves differently depending on whether the goal is to maximize SE or EE. In particular, binary transmit power control is generally optimal for maximizing SE, whereas a range of transmit power levels can be used to maximize EE. This distinction will later affect the performance of the LLM-based resource allocation.

\subsection{Proposed Resource Allocation based on LLM}

In the proposed LLM-based resource allocation strategy, the transmit power of each transmitter is determined based on the channel gain, $h_{i, j}$. In this approach, the channel gain serves as the input, and the transmit power is the output of the LLM, such that the channel gain is provided as a prompt, and the resource allocation can be derived from the text generated by the LLM. Unlike the conventional DNN-based approach, which requires a customized structure, a generic LLM can be used for resource allocation. The procedure of proposed LLM-based resource allocation is depicted in Fig. \ref{fig_dnn_model}.

In our work, we adopt a few-shot learning-based approach, where the channel gains and the corresponding transmit power strategies are provided as references in the prompt of LLM. Notably, we do not provide a formal description of the optimization objective; instead, we simply provide the channel gain and its corresponding optimal strategy. As a result, our framework can be applied to other optimization problems as well. 
In the generation of prompts for the LLM, we normalize the channel gain to have zero mean and unit variance, then multiply it by 100. This value is subsequently rounded to an integer. Similarly, the transmit power is normalized by its maximum value, i.e., $P_{\textrm{T}}$, and multiplied by 100, and is rounded to an integer. This preprocessing of numerical values is crucial because the LLM recognizes numeric values as strings, making it important to simplify these values for efficiency. Specifically, the original channel gain is typically on the order of $10^{-9}$, which is inefficient to express in string format. Additionally, given that the number of inputs to the LLM (i.e., the number of tokens) is limited, it is desirable to reduce the number of characters used to express each numeric value. For example, the formulated prompt to maximize the total SE can be illustrated as follows:

\textit{"If A is -29, 30, 128, -26, then B is 0, 100. If A is -31, -31, -19, -31, then B is 100, 0. $\cdots$ If A is 51, -32, -22, -19, then B is "}

In the illustrative example above, the channel gain is denoted as $A$ in the order of $\hat{h}_{1, 1}$, $\hat{h}_{1, 2}$, $\hat{h}_{2, 1}$, $\hat{h}_{2, 2}$, where $\hat{h}_{i, j}$ is the preprocessed channel gain $h_{i, j}$. On the other hand, the transmit power aimed at maximizing the total SE, is denoted as $B$ in the order of $\hat{P}_{1}$, $\hat{P}_{2}$, where $\hat{P}_{i}$ represents the preprocessed transmit power. In the considered few-shot learning approach, the training data includes the values of both $A$ and $B$. However, for the channel gain in the last line, where we need to determine the transmit power, the value of $B$ is omitted. Since the LLM is designed to generate text that follows the provided prompt, it will generate the appropriate resource allocation for the given channel gain. The first output of the LLM is then converted to the transmit power of each transmitter. Note that if the output of the LLM is not in the form $\hat{P}_{1}$, $\hat{P}_{2}$, we assume that the LLM fails to find the appropriate transmit power such that we set the transmit power for both transmitters to zero.

Unlike the conventional DL-based approach, where transmit power is always output as numeric values, the LLM-based approach requires converting string text to numeric values that must follow a predefined format. If the output does not follow this format, resource allocation cannot be determined properly. In addition, the LLM may generate completely nonsensical output. To mitigate these drawbacks, we considered the joint utilization of a simple binary transmit power control scheme \cite{Lee2021a}. Specifically, in addition to the LLM, we consider a simple binary transmit power control scheme, where either one transmitter operates at its maximum power. In our scenario, we need to compare only two cases: whether transmitter 1 transmits at maximum power or transmitter 2 does. We then compare the achievable rate of the LLM-based scheme with that of the binary scheme and select the strategy that provides higher performance. 

\subsection{Performance Evaluation}

In the performance evaluation, we assumed that the transmitters and receivers were randomly distributed over an area of 30 m $\times$ 30 m. We used the following parameters by default: $W =$ 10 MHz, $N_{0}$ = -173 dBm/Hz, $P_{\textrm{M}}$ = 20 dBm, and $P_{\textrm{C}}$ = 30 dBm. We considered a simplified path loss model with a path loss coefficient of $10^{3.453}$ and a path loss exponent of $3.8$. In addition, we used an independent and identically distributed circularly symmetric complex Gaussian random variable for multipath fading, with zero mean and unit variance.

For the LLM structure, we considered three different LLM models based on LLaMA, each fine-tuned on different datasets, namely CodeLLaMA-7B, LLaMA-2-7B-32K-Instruct, and LLaMA-2-7B. First, the CodeLLaMA-7B model, which we refer to as \textit{LLM 1} in our performance evaluation, is fine-tuned using source code, making it specialized for programming tasks and offering robust support for code generation and understanding. Second, the LLaMA-2-7B-32K-Instruct model, denoted as \textit{LLM 2} in our performance evaluation, is fine-tuned to follow complex instructions with a relatively long context length. This model is well-suited for tasks that require detailed execution of instructions. Third, the LLaMA-2-7B model, which we refer to as \textit{LLM 3}, is the base LLaMA 2 model that has not been fine-tuned for specific tasks. All considered models have 7 billion parameters, which are quantized to 5 bits to reduce the model size.

In the performance evaluation, we considered two proposed schemes. First, we considered the purely LLM-based scheme, which we refer to as \textit{Prop. 1}. This approach relies solely on LLM for resource allocation. Second, we evaluated a joint scheme that combines the LLM-based approach with binary transmit power control, which we refer to as \textit{Prop. 2}. In this scheme, the LLM is used together with a simple binary transmit power control method to improve the reliability and mitigate the shortcomings of the purely LLM-based approach. We also present the optimal performance obtained by exhaustive search as a benchmark. Furthermore, we include the performance of a random scheme, where the transmit power is randomly assigned, and a binary transmit power control scheme.

\begin{figure}[t]
	\centerline{\includegraphics[width=7.5cm]{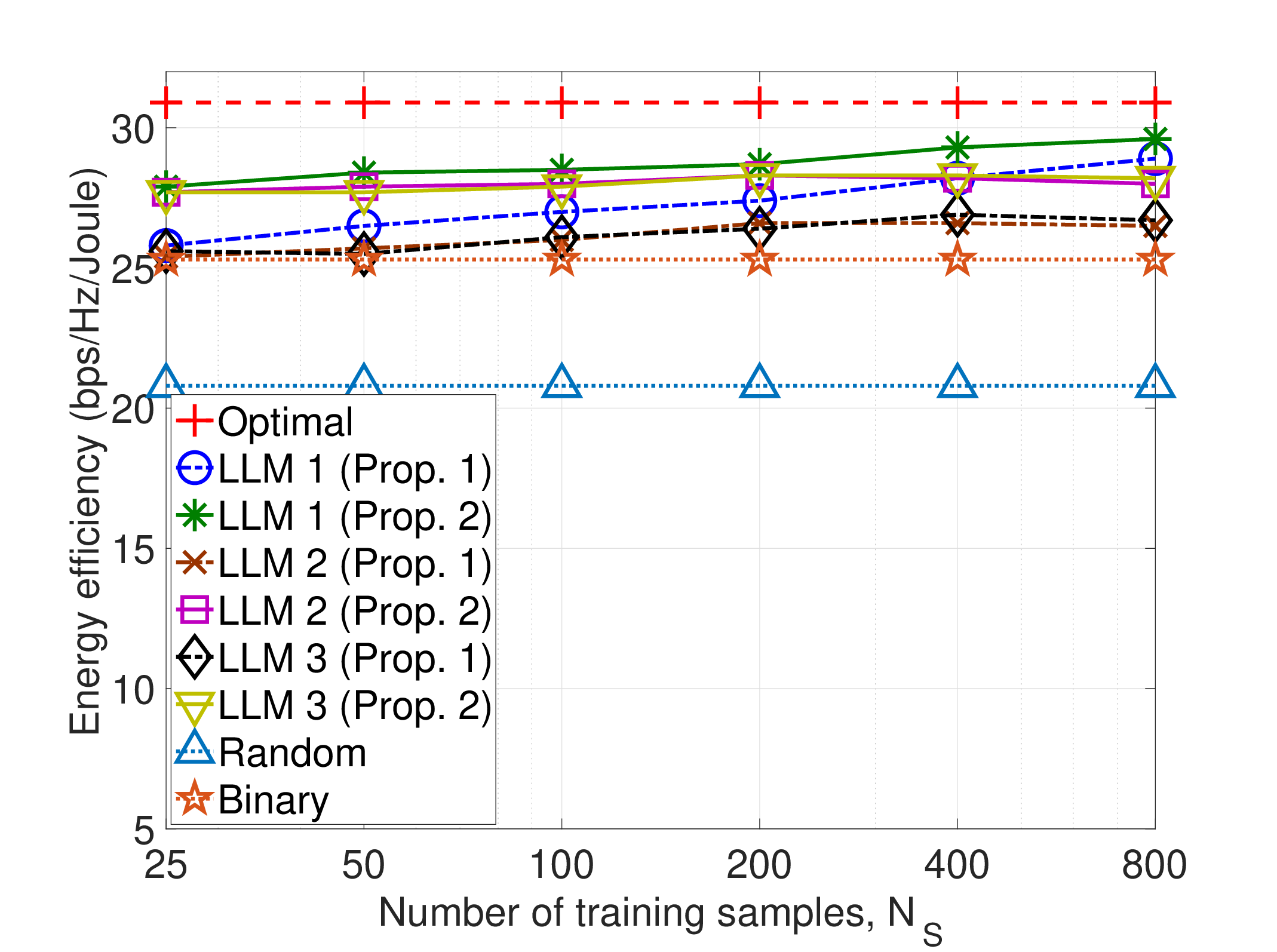}}
	\caption{EE vs. number of training samples.}
	\label{fig_EE_NS}
\end{figure}
First, in Fig. \ref{fig_EE_NS}, we show the achievable EE by varying the number of training samples fed through the prompts of the LLM, denoted as $N_{\textrm{S}}$. As the number of training samples increases, the achievable EE of the LLM-based approaches also increases, achieving up to 96\% of the optimal performance in the best case. Notably, even with only 25 training samples, the LLM-based approach outperforms the binary transmit power control scheme. In addition, we can find that the achievable EE of the LLM-based scheme is significantly higher than that of the random scheme, confirming the importance of proper resource allocation. By comparing the achievable EE of Prop. 1 (purely LLM-based) and Prop. 2 (joint LLM and binary transmit power control) for the same LLM models, we observe an improvement in EE of about 2-10\% in Prop. 2. This result validates the need to integrate conventional resource allocation strategies to improve the performance of LLM-based approaches. Furthermore, we can find that the performance varies significantly among the different LLM models used, i.e., LLM 1 (CodeLLaMA-7B) achieves about 10\% higher EE compared to the other two LLM models. We conjecture that the logical capabilities of the CodeLLaMA-7B model for the source code generation allow it to outperform the other LLM models.

\begin{figure}[t]
	\centerline{\includegraphics[width=7.5cm]{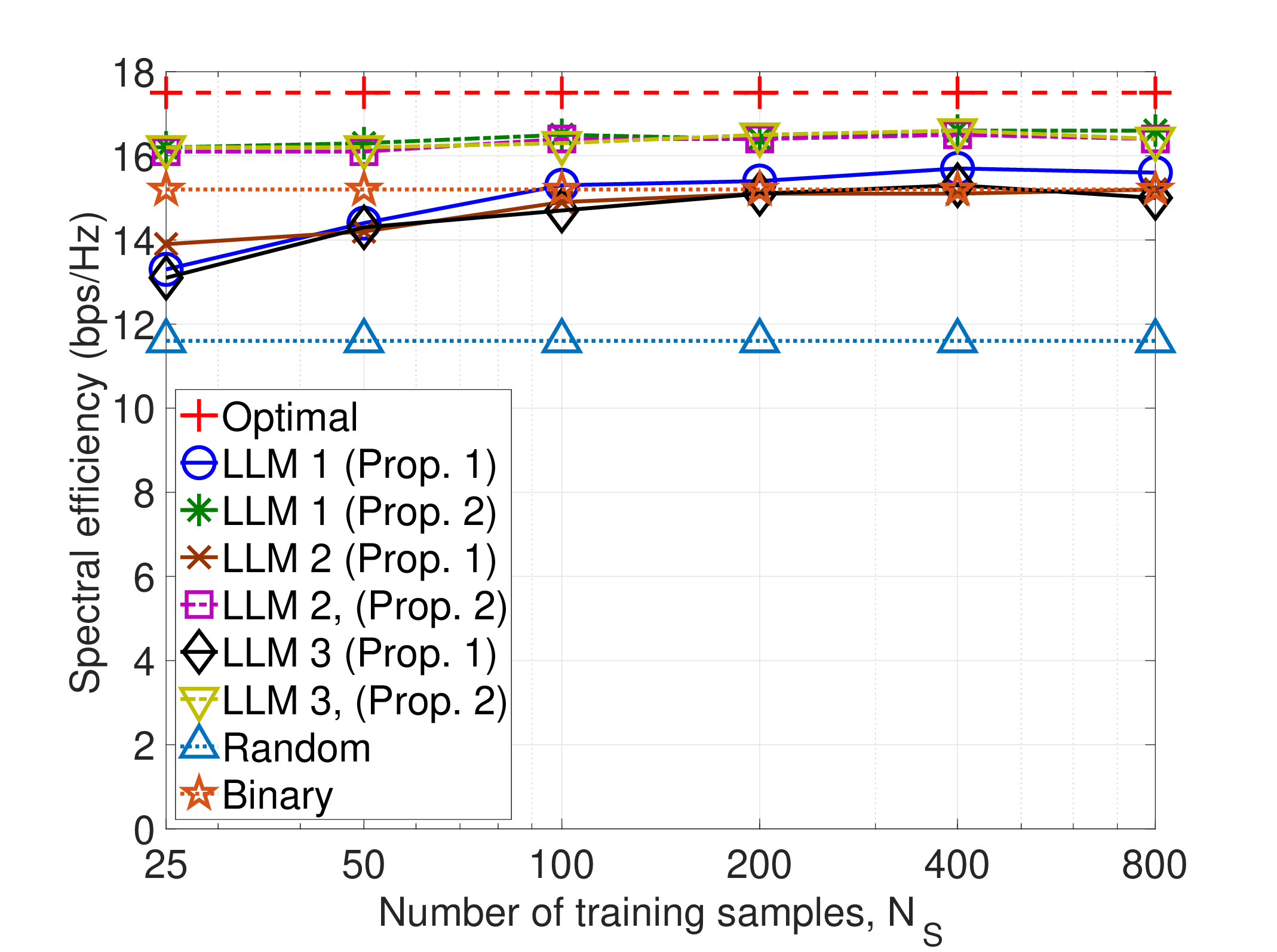}}
	\caption{SE vs. number of training samples.}
	\label{fig_SE_NS}
\end{figure}
In Fig. \ref{fig_SE_NS}, we show the achievable SE as the number of training samples is varied. As expected, the achievable SE increases with the number of training samples. However, compared to EE maximization, the performance gains of the LLM-based approach for SE are relatively small. Specifically, the SE achieved by Prop. 1 (purely LLM-based) is only 3\% higher than that of the binary transmit power control scheme, achieving 94.8\% of the optimal performance in the best case. We conjecture that this is because the optimal resource allocation strategy for SE is close to the binary transmit power control scheme, which involves abrupt changes in transmit power levels and such abrupt changes may pose a challenge for the LLM to predict accurately. Furthermore, similar to the results in Fig. \ref{fig_EE_NS}, LLM 1 (CodeLLaMA-7B) outperforms the other LLM models. Additionally, the joint utilization of conventional schemes alongside the LLM-based approach, i.e., Prop. 2, improves the achievable SE by 5-10\%.

\begin{figure}[t]
	\centerline{\includegraphics[width=7.5cm]{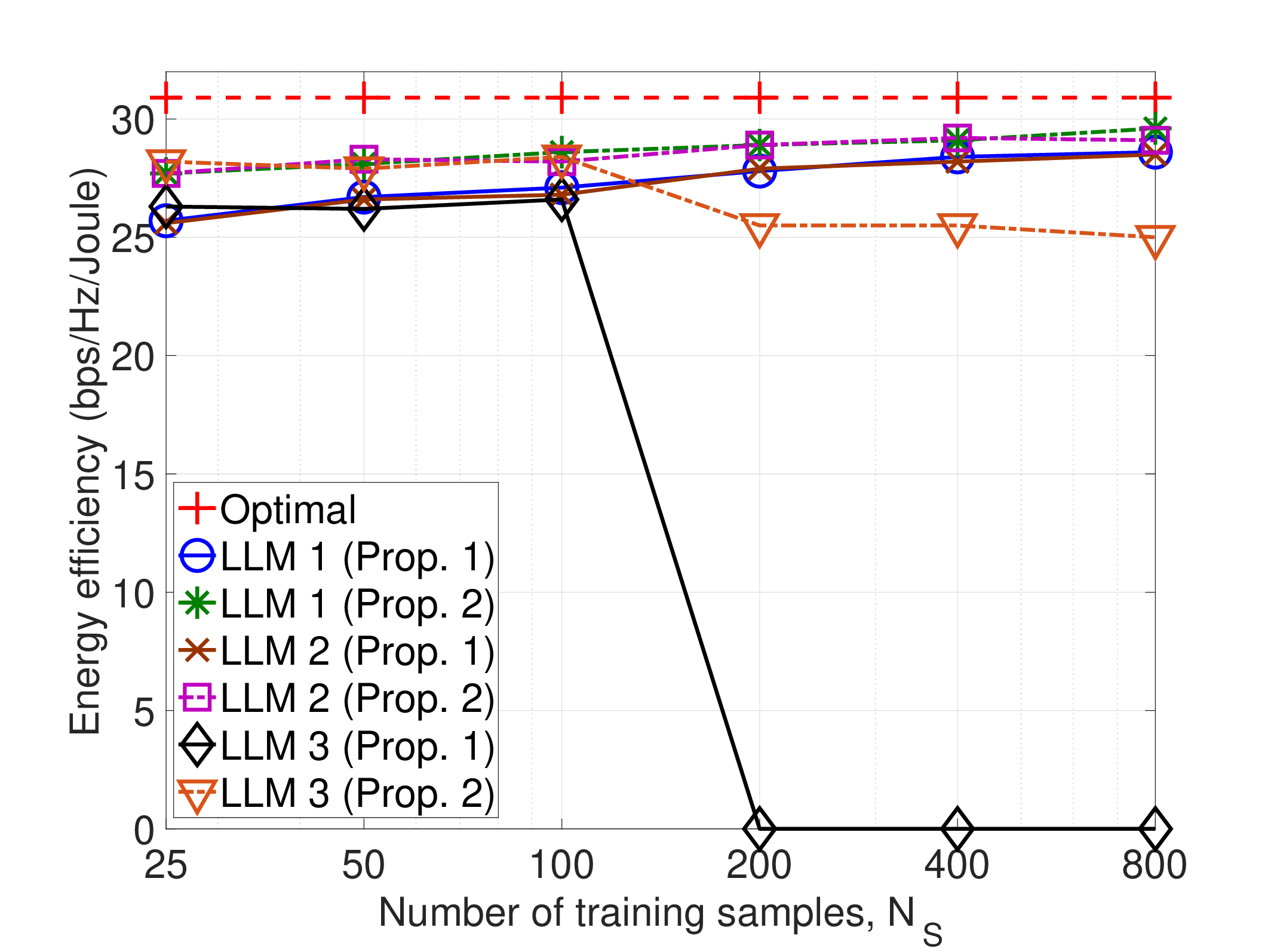}}
	\caption{EE vs. number of training samples for larger LLM models.}
	\label{fig_EE_BIG}
\end{figure}
In Fig. \ref{fig_EE_BIG}, we show the achievable EE as the number of training samples is varied for larger LLM models, specifically those with 13 billion parameters. For clarity, we have omitted the performance data for the random scheme and the binary transmit power control. We observe that the overall trend is similar to that in Fig. \ref{fig_EE_NS}, with EE improving as the number of training samples increases. However, the performance improvement gained from using larger LLM models is relatively small, with an increase of less than 1\%. This suggests that the use of large LLM models may not be necessary for resource allocation tasks, which is beneficial for communication systems where the use of large LLMs is challenging. Notably, we find that LLM 3 (the base LLaMA 2) fails to generate a viable resource allocation strategy when the number of training samples exceeds 100. During the experiment, we discovered that the LLM 3 begins to generate output in a different format in these cases, highlighting the importance of integrating conventional schemes that can compensate for the potential malfunctions or inconsistencies of LLM-based approaches.

\begin{figure}[t]
	\centerline{\includegraphics[width=7.5cm]{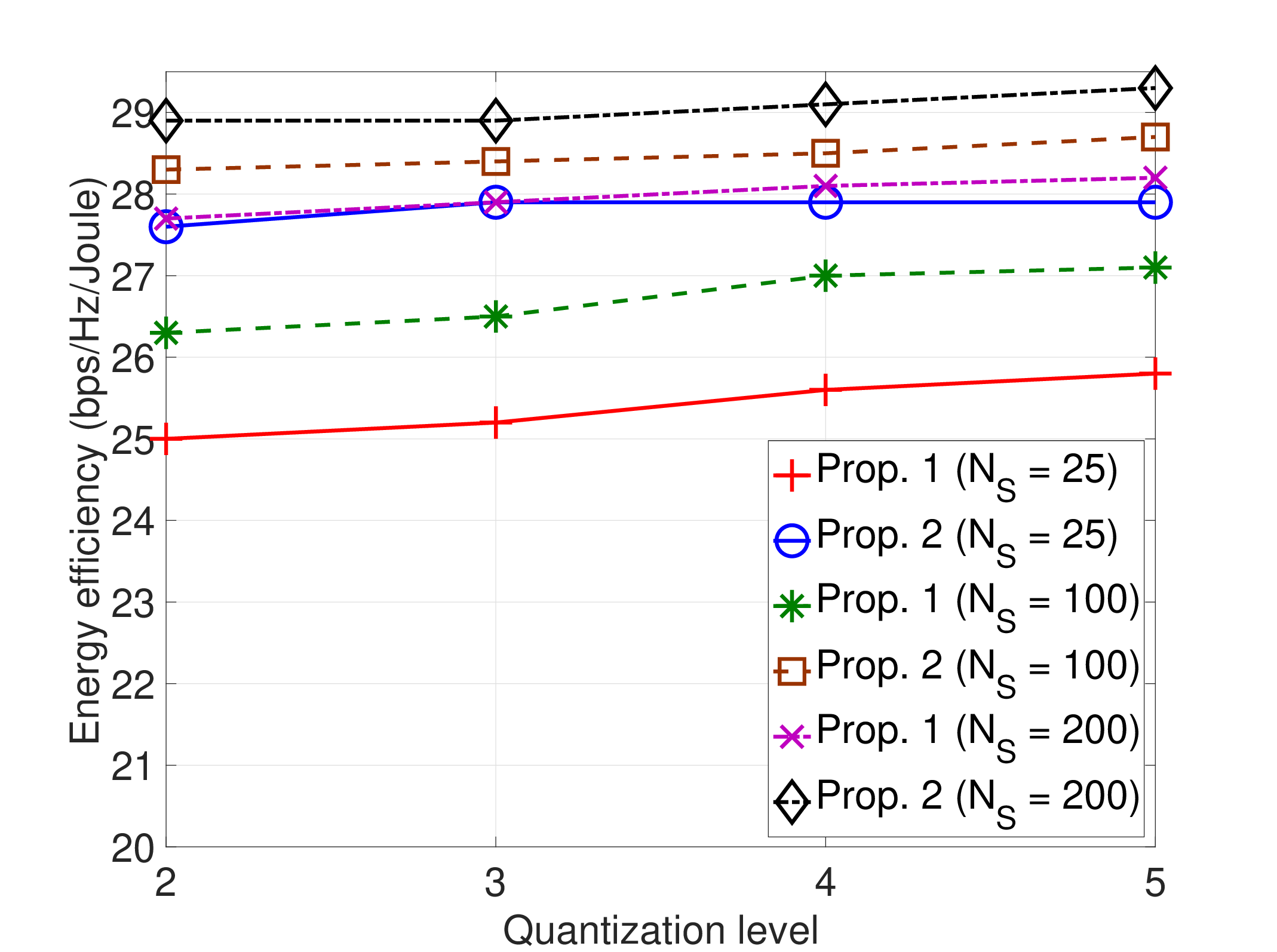}}
	\caption{EE vs. quantization level.}
	\label{fig_EE_QA}
\end{figure}
Finally, in Fig. \ref{fig_EE_QA}, we compare the performance of the LLM-based approach across different quantization levels of the LLM. It is important to note that quantizing parameters in LLM models is a common practice due to the considerable size of unquantized models. For simplicity, only LLM 1 (CodeLLaMA-7B) was considered in the simulation. As expected, the achievable EE improves with higher quantization levels, although the improvement is relatively small. In particular, the performance improvement is more significant when the number of training samples is small, showing a 3.2\% improvement in EE. This improvement decreases as the number of training samples increases, with a 1.8\% improvement observed when the sample size reaches 200. This result suggests that smaller LLMs can be effectively used for resource allocation tasks, indicating promising potential for the use of LLMs in resource allocation.

\section{Research Challenges} 

In the following, we discuss the research challenges associated with the future use of LLMs for resource allocation.

\subsection{Latency and Computation Time}

In general, the size of LLMs is extremely large due to their large number of parameters. GPT-4, for example, has 1.7 trillion parameters. As a result, running LLMs requires significant computation resources, often necessitating the use of cloud-based services, which can introduce long latency. In addition, due to the size of LLMs, the computation time required can be large. To address the issues of long latency and computation time, the development of small LLMs (sLLMs) tailored for resource allocation is essential. These sLLMs, combined with edge AI technology, can distribute the computational load across distributed nodes, thereby reducing latency and improving efficiency.

\subsection{Optimized LLM Architecture}

To achieve higher performance, it is necessary to optimize the LLM architecture, including the proper selection of a pre-trained LLM model, which requires extensive performance evaluation against different LLM models. Moreover, diverse multimodal input data, such as user mobility, can also be utilized. For example, proactive resource allocation can be achieved by predicting user activity. Additionally, the energy consumed during the execution of LLMs should be carefully considered, as they utilize substantial computing power to determine the resource allocation strategy.

\subsection{Training Methodology}

In our illustrative example, we employed a supervised learning-based approach, that utilizes multiple labeled resource allocations for different channel gains in a few-shot learning framework. However, the acquisition of such labeled data is often challenging in practice, highlighting the need for innovative resource allocation techniques capable of finding optimal strategies without relying on labeled data. This capability is also critical for ensuring the scalability of LLM-based scheme, as the number of tokens (i.e., the input size to the LLM) cannot be arbitrarily large. To develop more effective training methodologies, careful prompt engineering and fine-tuning of LLM should be considered.

\subsection{Interpretability and Explainability}

Similar to DL-based approaches, the output of LLMs can be unpredictable, potentially resulting in non-optimal performance. In particular, unlike DL-based methods that output numerical values, LLMs generate text-based outputs, which can be difficult to handle. Furthermore, the performance of resource allocation derived from LLMs cannot be guaranteed because the operation of LLMs is considered as a black box. In addition, the hallucination of LLM-based approach can make the output unreliable. To address these issues, it is necessary to incorporate Explainable AI (XAI) techniques into LLMs and appropriately integrate conventional resource allocation methods, as suggested in the previous section.

\section{Conclusions}

In this article, we discussed the application of LLM for resource allocation. Unlike conventional DL-based approaches, the LLM-based approach eliminates the need to build and train dedicated DNNs for specific tasks. We considered a simple resource allocation problem and demonstrated the ability of LLMs for the resource allocation strategy. To address the drawbacks of the LLM-based scheme, we also devised a joint consideration of conventional resource allocation strategies. Through performance evaluation, we confirmed that the LLM-based resource allocation can achieve near-optimal performance and investigated its characteristics. In addition, we outlined challenges associated with the application of LLMs for resource allocation.

\begin{IEEEbiographynophoto}{Woongsup Lee}
[S'07, M'13] (woongsup.lee@yonsei.ac.kr) is currently an Associate Professor with the Graduate School of Information, Yonsei University, Seoul, South Korea. His research interests include large language model, deep learning, cognitive radio, future wireless communication systems, and smart grid systems.
\end{IEEEbiographynophoto}

\begin{IEEEbiographynophoto}{Jeonghun Park} [M'17] (jhpark@yonsei.ac.kr) received the Ph.D. degree from The University of Texas at Austin, USA, in 2017. He worked as a Senior Engineer at Qualcomm, USA. He is currently an Assistant Professor with the School of Electrical and Electronic Engineering, Yonsei University, South Korea.
\end{IEEEbiographynophoto}

\end{document}